# Conceptual Modeling of Inventory Management Processes as a Thinging Machine

Sabah Al-Fedaghi

Computer Engineering Department
Kuwait University
Kuwait
sabah.alfedaghi@ku.edu.kw

Nourah Al-Huwais

Information Technology Department
Kuwait National Petroleum Company
Kuwait
n_huwais@knpc.com

*Abstract*—**A control model is typically classified into three forms: conceptual, mathematical and simulation (computer). This paper analyzes a conceptual modeling application with respect to an inventory management system. Today, most organizations utilize computer systems for inventory control that provide protection when interruptions or breakdowns occur within work processes. Modeling the inventory processes is an active area of research that utilizes many diagrammatic techniques, including data flow diagrams, Universal Modeling Language (UML) diagrams and Integration DEFinition (IDEF). We claim that current conceptual modeling frameworks lack uniform notions and have inability to appeal to designers and analysts. We propose modeling an inventory system as an abstract machine, called a Thinging Machine (TM), with five operations: creation, processing, receiving, releasing and transferring. The paper provides side-by-side contrasts of some existing examples of conceptual modeling methodologies that apply to TM. Additionally, TM is applied in a case study of an actual inventory system that uses IBM Maximo. The resulting conceptual depictions point to the viability of FM as a valuable tool for developing a high-level representation of inventory processes.**

## I. INTRODUCTION

In general, inventory is defined as items stocked in a store to meet anticipated requests. Inventory management or control is a system to balance product needs with demand to minimize costs that arise from obtaining and holding inventory [1]. There are several schools of thought that view inventory and its functions differently. This paper presents a foundation for inventory processes modeling that views an inventory as an abstract machine, called a Thinging Machine (TM), with five operations that may include infrastructure of submachines. We claim that such modeling, which is based on TM, facilitates understanding and serves as a base for consequent phases of an inventory system's design and development. A model is understood as an abstract view of a portion of reality that enables developers to concentrate on relevant aspects of the system and discount needless complications [2].

Inventory management models are typically classified into three forms [3]:

- A conceptual model that contains text, pictures and diagrams to explain the terms and principles of a particular system's functioning.
- An analytical (purely mathematical) model that uses mathematical concepts and language and contains formulas for analysis and calculations.
- A simulation (computer) model that attempts to simulate an abstract model of a particular system.

The last two modeling techniques are concerned with minimizing the total cost of inventory based on a decision-making process considering the cost of holding the stock, placing an order or encountering a shortage (e.g., insufficient stock to meet demand). This paper is focused on *conceptual* modeling of inventory management systems.

Conceptual modeling pertains to identifying, analyzing and describing the essential concepts and constraints of a domain with the help of (diagrammatic) modeling language that is based on a small set of basic meta-concepts [4]. It helps in understanding and communicating among the stakeholders and serves as a base for consequent phases of a system's development [5]. It should reflect the reality of the organization and its operations; conceptual models are most valued in terms of completeness, faithfulness to the realization of the underlying real system, understandability and susceptibility analysis.

### A. Inventory management

In an inventory management system, several basic notions are recognized, including minimum and maximum stock level, safety and reorder points and timing. The basic function of a management system requires preserving items' quantities and maintains them in such a manner that they do not remain in stock for a long time, which would result in efforts waste (e.g., delays in production, work and maintenance). Inventory cost remains low when the correct quantities of products are ordered at the right price and time.

Control is established by fixing the minimum and maximum levels of stock. These levels are calculated based on historical data, the expected requirements and in view of the inventory's condition. Reaching the minimum level may result in stock running out; reordering occurs in such a way that items are received before the stock volume reaches the minimum level.

Minimum stock occurs when the stock falls below an established critical level at which the enterprise processes may be harmed. In such an event, a warning is sent to management to exert further efforts or extra resources to ensure that the situation is rectified. The maximum point is utilized to avoid any superfluous stocks that may result in halting the flow of items. Designated stock is maintained for safety considerations, especially for events such as a serious stoppage of operations or to maintain the reliability of supply. In most cases, it is equal to minimum stock. The reordering process ensures that items are not out of stock by taking into consideration current stock, lead time and receiving time before reaching a minimum level.

Whatever the adopted inventory system, an organization needs a conceptual description that describes its real-world domain and does not include any information technology aspects. It would be able to serve various levels of granularity and complexity, such as by serving as a guide for the subsequent information systems specification, analysis, design and validation.

### B. Problem and Solution

This paper claims that current conceptual models of inventory management systems lack comprehensiveness and completeness. Additionally, a lack of conceptual representation of processes makes it more difficult for end users to understand an existing process or simulate a new one. Accordingly, a high-level conceptual language can contribute by filling some needs and acting as a foundation in this area of research. As a step in this direction, the paper applies the recently developed TM that is based on the *thinging machine* notion and presents a different conceptualization of such. This paper advances the inventory management processes in a holistic way by developing a framework that is sufficiently inclusive. Generally, TM provides a diagrammatic representation of the static, dynamic and control specifications at play in the inventory management processes.

To show the viability of the proposed methodology, we use TM in an actual case study of an inventory management system that is currently being implemented using IBM software without an explicitly documented conceptual description.

### C. Appraoch

Apparently, it is very difficult to contrast the involved diagrammatic models because they are based, to a large extent, on factors such as understandability (that pertains to visualization and graph completeness). A straightforward way to accomplish that is to put different diagrammatic depictions side-by-side and judge them accordingly. Therefore, we will give a few examples of current techniques so that, at the end, we are able to observe and contrast different samples.

### D. Sample diagrams

Today, most companies utilize computer systems for inventory control in which withdrawals are recorded and the inventory balance is monitored. Orders are placed and the balance of stock is updated by the computer. Modeling inventory processes is an active area of research that uses many methods, including flow charts, data flow diagrams, Universal Modeling Language (UML) diagrams, role interaction diagrams, Gant charts and Integration DEFinition (IDEF) [6].

In this section, several samples of inventory-related diagrams are presented. The purpose is not intended to give a fair discussion of these examples; rather, the aim is to provide an awareness of the type and nature of conception and depiction upon which this method is built. The samples will also provide the opportunity to contrast the diagramming techniques after presenting TM diagrams of our case study.

Saraste [7] used the flowchart technique, which is "very flexible and easy to use" [7], as shown in Fig. 1. In the inventory control environment, the modeling used by Saraste [7] may not be suitable to model the system in a holistic way through developing a framework that is sufficiently inclusive. In certain situations, the entire enterprise processes may be harmed by local event (e.g., stock goes below an established critical level). Having knowledge of the entire view of the system—something that is missing in flowcharting—can help management employ efforts and/or resources to prevent an adverse effect on a production situation. Another flowchart-based representation is a sample of the Maximo diagram in an IBM Knowledge Center [8], as partially shown in Fig. 2. Flowcharts were the target of many criticisms regarding their value in design and education [9-10] that have nearly led to their elimination. Lately, they have also been revived in the form of a UML activity diagram. According to Storrle and Hausmann [11], "activity diagrams have always been poorly integrated, lacked expressiveness, and did not have an adequate semantics in UML."

According to Patel [12], Entity Relationship (ER) modeling allows for the formation of high-level conceptual data models that can be used to form a graphical representation for design, as seen in the initial ER diagram in Fig. 3.

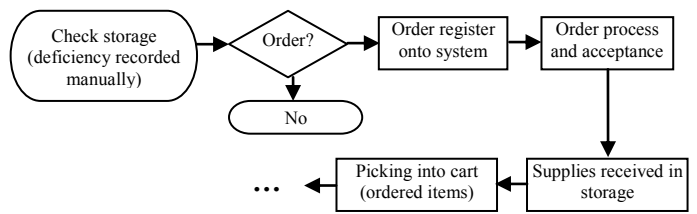

**Fig. 1 Order modeling using a flowchart (partially redrawn from [7])**

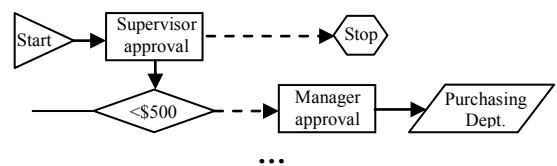

**Fig. 2 IBM Knowledge Center diagram that involves a flowchart (partially redrawn from [8])**

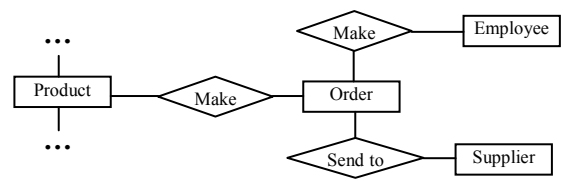

**Fig. 3 ER diagram (partially redrawn from [12])**

Nevertheless, difficulties in ER modeling (e.g., temporal aspects) are well known [13]. Rinardwiatma [14] used workflow process diagrams, such as the one shown in Fig. 4, to integrate inventory management in an IBM Maximo-based system.

Most existing conceptual modeling methods use object-oriented methodology that employs UML as a foundation that requires breaking the system structure and behavior into several types of diagrams, then further decomposing them into other diagrams. It is claimed that this approach has many advantages, such as simulating the modeler's way of thinking [15] and contributing to the reduction of complexity in the representation of technical systems and design processes [16]. For example, Tchantchane [17] utilized UML use case, class, state and sequence diagrams in designing a sales ordering system (as shown in Figs. 5-8).

Nevertheless, According to Mordecai [18], there is a "significant inability of common conceptual modeling frameworks to appeal to practicing designers and analysts." These diagrams are completely heterogeneous with several different conceptual bases. The purpose of this heterogeneity is to achieve a wide range of options for expression, depending on the situation. Shoval and Kabeli [19] suggested a merger of the data flow diagram, the ER diagram and object-oriented constructs. The multiplicity of diagrams for the same dilemma in UML is a known problem [20] that contrasts with providing a single, integrated diagrammatic representation that incorporates function, structure and behavior.

The next section introduces TM to be used both as a thinking style and as a vehicle for depicting the capturing inventory processes [21-26].

## II. THINGING MACHINE (TM)

Reality consists of a range of things, such as an externally experienced object, situation, event or action, or a privately experienced sensation, mood, emotion, memory or thought. These things are the content of our *wajood* (existence). Some of these things comprise others or they form an environment or a place for others.

The thing tree takes the thing carbon dioxide from the thing air and gives it the thing oxygen. We say that the tree *receives* carbon dioxide from the air and *releases* and *transfers* oxygen to it. Also, the thing tree *processes* carbon dioxide to *create* oxygen. The tree in this case is a *machine* that releases, transfers, receives, processes and creates things. By the same conceptual manner, the thing human being is a machine that receives the thing oxygen and processes it to create the thing carbon dioxide that is released and transferred to the atmosphere. These things/machines are called TMs. The machine can be conceptualized as a *lived atmosphere* (i.e., environment/space/capsule) of its things where they are created, processed, received, released or transferred. We describe a TM diagrammatically as shown in Fig. 9. The figure is a generalization of the typical input-process-output model that is used in many scientific fields. The machine is constructed from the sub-machines of flows including the machine itself.

Martin Heidegger [27] describes viewing a particular tree as, in our model, a machine. It is rooted in the earth with its trunk rising up and branches splayed out, swaying in the wind.

It is inhabited by many tiny creatures, and it responds to the wind currents. The tree is a certain compilation of the threads of life. It is a thing when we see it as a glimpse of life-information, never the same from one moment to the next [28]. It is not only the tree that is a machine, but also people, animals, towns, the sun and clouds, as well as day, night, feelings, numbers, atoms, data, etc.

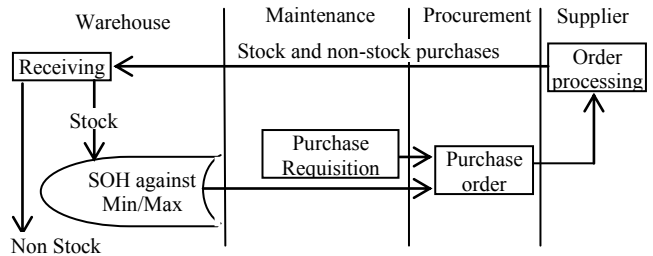

Fig. 4 Inventory flow process (partially redrawn from [14])

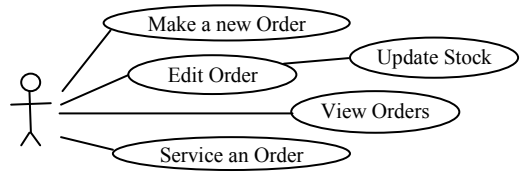

Fig. 5 Use case diagram (partially redrawn from [17])

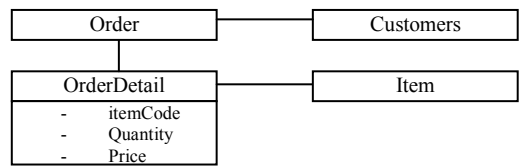

Fig. 6 Class diagram (partially redrawn from [17])

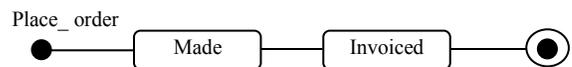

Fig. 7 State diagram (partially redrawn from [17])

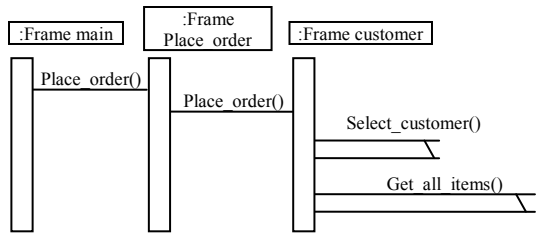

Fig. 8 Sequence diagram (partially redrawn from [17])

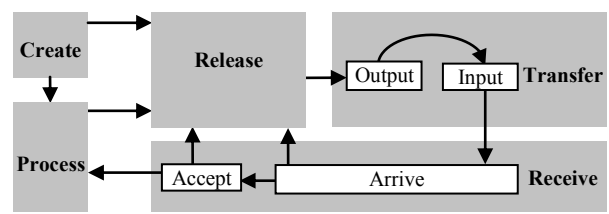

Fig. 9 Thinging Machine

The (unproven) claim in TM modeling is that the five operations—create, process, release, transfer and receive—are sufficient to represent all activities in a machine. The only justification for this is the diverse modeling of many systems that appear in many publications, including the modeling inventories in this paper.

*Things* that flow in TM refer to the exclusive (i.e., there is one and only one) conceptual movement among the five operations (stages) shown in Fig. 10. It may be argued that *things* (e.g., goods) can also be *stored* in addition to being created, processed, released, transferred and received. However, because *stored* is not a generic operation, things can be stored after being created, hence becoming *stored created data*, or, after being processed, becoming *stored processed data* and so on. When all arriving things are accepted, then arrive and accept are represented by receive.

*Create* (emerge) is one starting point of a thing in a machine, in addition to being imported (transfer/receive) from the outside. Through creation and importing the system (machine) becomes aware (recognition) of a new thing in the machine. Additionally, a thing "disappears" from the "radar" of a machine, either when it is *de-created* (e.g., deleted) or by departing (release/transfer). Note that a thing can be released, but not transferred (e.g., finished goods waiting to be shipped when a truck arrives) or a thing being transferred (input), yet not arriving (e.g., an email arriving, but an error preventing the recipient from accessing it).

*Process* means that the machine changes the thing in a certain way. For example, a doctor machine processes a patient to decide how to treat him/her.

Each type of flow of things is distinguished and separated from other flows. No two streams of flow are mixed, just as lines of electricity and water are separated in buildings' blueprints. However, two types of things can enter a machine of a super type of thing (e.g., integers and real numbers flow to a number machine). A TM does not need to include all of the stages; for example, an archiving system might only use the stages transfer, receive, release and process (i.e., not create).

Multiple machines can interact with each other through flows or by triggering stages. Triggering is a transformation (denoted by a dashed arrow) from one flow to another (e.g., a flow of electricity triggers a flow of air).

## III. Inventory in IBM Maximo

This section applies TM to model an actual inventory system. The system extends over physical and digital spheres, where it flows across different machines changing their forms. This paper assesses a case study that uses IBM Maximo, an asset management software system. In Maximo, nodes can represent various points in a business process (e.g., start node, condition node, interaction node, sub-process node, task node and stop node). A workflow process is created by interweaving nodes and connection lines within the workflow. There are many notions: person records, role records, action records, communication templates, notifications, escalations and action groups, etc. Also, there are many actions (e.g., create, change, incident, problem, service request and work order) [29].

In our case study, Maximo modules are used to manage inventory, including functions such as controlling inventory, making purchases and tracking stock levels and contracts.

Inventory is a central module in Maximo. It functions in a dynamic relationship with the preventive maintenance, work orders, contracts, purchasing and assets modules. Maximo can automate processes that are repetitive or occur at regular intervals; for example, preventive maintenance, periodic inspections or reordering inventory items" [30].

In Maximo, inventory management is an important part of maintaining any asset for which the inventory module in Maximo tracks the required materials (e.g., monitoring reorder points, purchase requisitions and purchase orders; tracking the movement of items into and out of inventory; issuing work orders; etc.).

TM and Maximo have a completely different conceptual view. Maximo-based systems utilize different types of diagrams to document and describe various applications, including inventory management systems. Assuming knowledge of basic Maximo terms, consider a difference in terms of Maximo *location* and *asset* (e.g., a bank as the location). From the standpoint of IBM Maximo, a location is a physical place, an operating position where equipment resides [31]. It is a place where assets are operated, stored or repaired. Fig. 10 contrasts the conceptualization of a location and an asset. In TM, the location is a machine while the asset is a submachine and a thing that flows.

In TM, an inventory system is a machine that serves other machines by managing their input/output flows so that they are near steady states. Ideally, flows in its stewarded machines are never stopped or slowed due to a lack of item supply. The machine also ensures that the supply does not clog them with overstocked items.

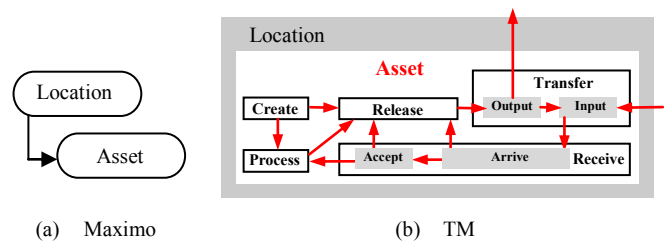

(a) Maximo          (b) TM

**Fig. 10 Maximo and TM representations of *location* and *asset* (The left part is a partially redrawn part of a figure [31])**

## IV. Case Study

The TM representation is applied to the process of responding to an approved requested order arriving from its department to the commercial department. Accordingly, the commercial department checks the inventory for the request. Three possible cases are based on the inventory status: the requested items are available, partially available or not available. These cases can be modeled using a TM to produce TMs that involve the inventory system (as shown in Fig. 11).

### A. A request arrives and current inventory is checked

Fig. 12 describes the general process of this case study. It is expanded into three cases that are explained in the following sub-sections. As shown in the figure, an approved request of an item (circle 1) from the requesting department (2) flows (3) to the commercial department system (4) to check the inventory control. The commercial department system, in this case study, is responsible for managing the inventory and ordering items.

Once the approved request is received, the commercial department triggers (5) a check of the inventory status (6) in the inventory machine (7) to examine the current stock (8) of available items that pertain to the request, which is a global variable and is initially set to zero. The current stock is processed and compared to the minimum level of the inventory (9). The minimum level is considered a point of emergency (i.e., when a certain item volume reaches a level that is considered below critical). The availability cases based on this comparison are:

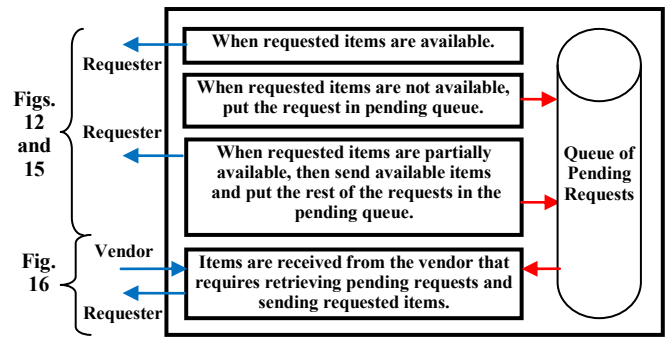

Fig. 11 Different portions of the inventory system

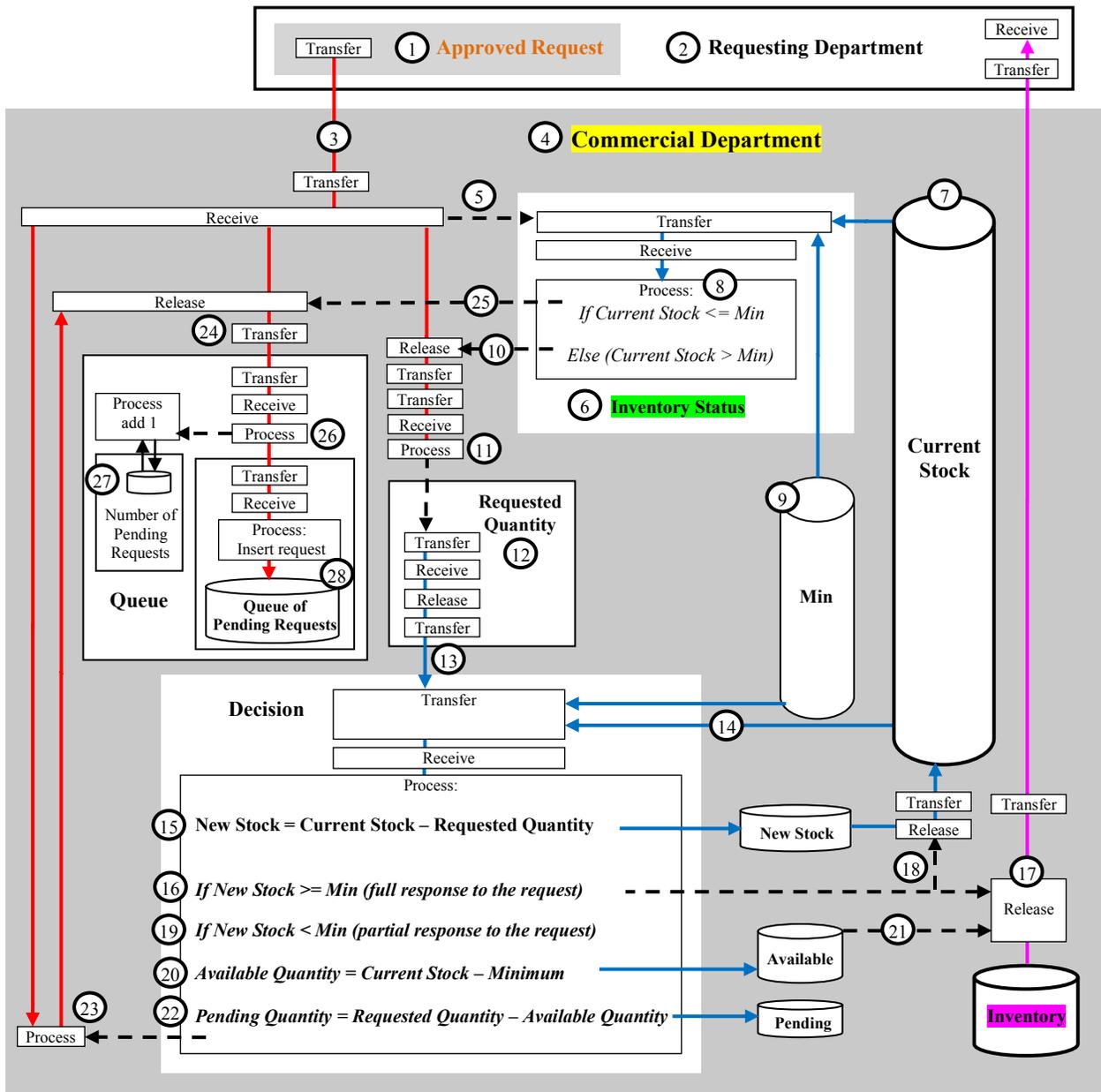

Fig. 12 General TM representation of the inventory control case study

- **If the current stock of the request is above the minimum level**: The request is processed (11) to extract (12) the requested quantity that flows to a decision machine (13), which also receives the current stock value (14). In the decision machine, the new stock is calculated as the current stock minus the requested quantity (15). The new stock is considered a global variable and is initially set to zero.
- **Accordingly,**
  (i) If the new stock is equal or above minimum (16), meaning that there are enough items to be delivered, then the decision machine triggers the release of the requested items (17) to be directly delivered to the requesting department and the replacement of the current stock by the new stock (18).
  (ii) If the new stock is less than the minimum (19), then a partial response to the request is possible:
    - The available quantity of items is calculated (20) as (the current stock minus the minimum) and released (21) to the requesting department (assuming it accepts that).
    - The pending quantity is calculated as the requested quantity minus the available quantity (22). The request is processed (23) to make it an inquiry for pending quantity and is sent to the queue system (24).
- **If the current stock of the request is equal to, or less than, the minimum level**: The request is sent to the queue system (26), where it is processed to increment the number of queued requests (27) and to add the request to the queue (28).

### B. Modeling events

Fig. 11 in the previous sub-section reflects a static structure of distributing orders among the three cases. To model the dynamic behavior of the case when a request arrives and the current inventory is processed, we need the notion of a machine *being-in-time*. In being-in-time, the machine not only creates, processes, receives, releases and/or transfers, but it also *machines* (habitually does these operations again and again). Time in TM is a *thing* that can be created, processed, released, transferred and received. Consider *the request has arrived and received in the commercial department*, which can be represented as shown in Fig. 13.

An event is a machine in a TM that contains at least three submachines: the time, the region and the event itself. The region is where the event *takes place*. The event in Fig. 13 includes the three machines: the region of the event (circle 1), which is a subdiagram of Fig. 12; the (real) time submachine (2); and the event submachine itself (3/green). It was previously stated in section two that the machine is constructed from the submachines of flows, including the *machine itself*. The machine itself is distinct from all of the submachines within. For simplicity's sake, an event will be represented only by its region.

Accordingly, we can identify the events in Fig. 14 as follows.

Event 1 ($E_1$): A request is received.
Event 2 ($E_2$): The inventory status is checked.
Event 3 ($E_3$): The current stock exceeds the minimum.
Event 4 ($E_4$): The request is sent to the decision procedure.

Event 5 ($E_5$): The new stock is calculated.
Event 6 ($E_6$): The new stock is = or > than the minimum.
Event 7 ($E_7$): Items are sent to the requester.
Event 8 ($E_8$): The new stock is less than the minimum.
Event 9 ($E_9$): The available and pending quantities are calculated, and the request is modified and sent to the queue system.
Event 10 ($E_{10}$): The quantity (circle 25 in Fig. 12) is equal to or less than the minimum.
Event 11 (E11): The request is added to the queue system.

Fig. 15 shows the chronology of these events. It can be used in the execution and control (event operations) of events as exemplified in the figure. While a machine machines, control is an awareness of this machining that creates second-level machining.

### C. If new stock for the request is below minimum

Fig. 16 models the machine in which the new calculated stock is less than the minimum (circle 19 in Fig. 12). In this case, as shown in Fig. 16, the quantity of items available is insufficient to be delivered to the requestor. Therefore, the request is partially satisfied by sending the available stock (the current stock minus the minimum) to the requestor (circles 1 and 2). Moreover, the updated current stock (3), which is now equal to the minimum, and the number of pending items (the requested quantity minus available quantity (4) are sent to the supervisor to decide upon making new supply order (5).

The supervisor applies the company's inventory policy (6), which is based on statistics that pertain to the ordering level, the minimum level and the maximum level. The maximum level is set according to the average of historical data to maintain the number of items typically on hand in recent years. These levels are determined by the commercial manager.

The supervisor (7) decides whether to issue a request for quotation (RFQ) (8). Moreover, the supervisor assigns (8) an employee from the commercial department as the declared buyer (9 –bottom right of the figure) who is tasked with the responsibility of the transaction for this RFQ. The RFQ flows to the team leader (10) to be further processed.

- If the requesting supervisor does not have the authority for such a type of RFQ (11), then it is canceled (12) by the team leader and a cancellation note is created and sent to the supervisor.
- If the specification of the RFQ is incorrect (13), then it is rejected and a rejection note is sent (14) to the supervisor in order to modify the RFQ specifications. Once the RFQ is modified (15), it is sent back (16) to the team leader.

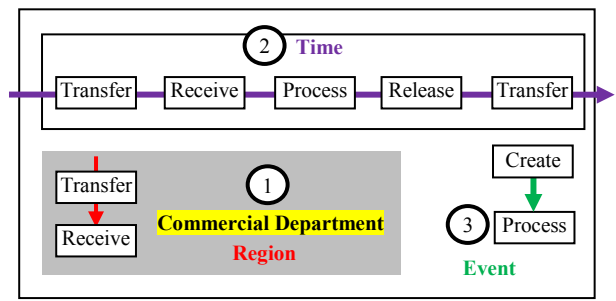

**Fig. 13 The event: *The request has arrived and received in the commercial department***

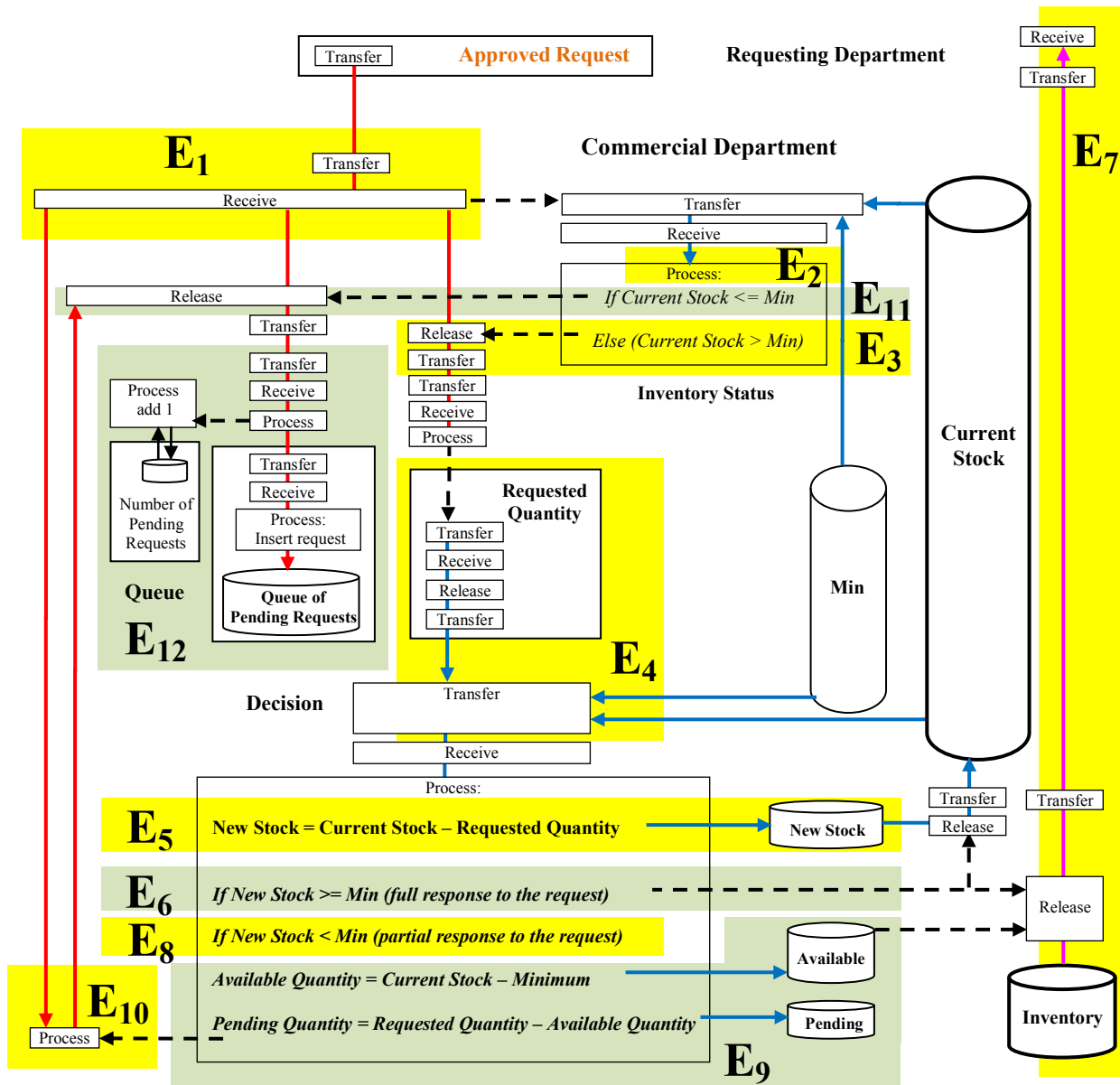

**Fig. 14 Events of the FM representation of the inventory control case study**

If the specifications are still incorrect (17), then another rejection process is repeated (18). Otherwise, it is approved (19).

- If the RFQ is neither cancelled nor rejected (20), then it is approved (21) by the team leader. Approvals flow to the supervisor (22 - copy) and to the manager (23) for further processing.

- The same cycle of the team leader's actions is repeated for the manager (24) and its description is omitted here. Assuming that the manager approves the RFQ, the approval is sent to the supervisor (25 – copy) and to the declared buyer (26) who was specified by the supervisor.

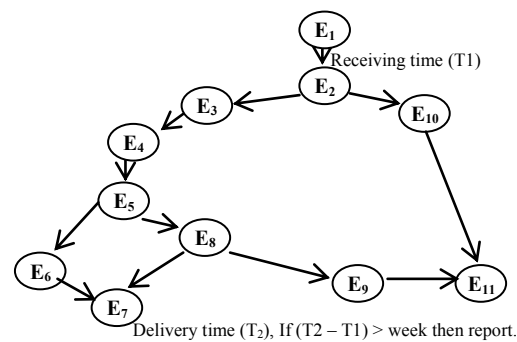

**Fig. 15 The chronology of events of TM representation of the inventory control case**

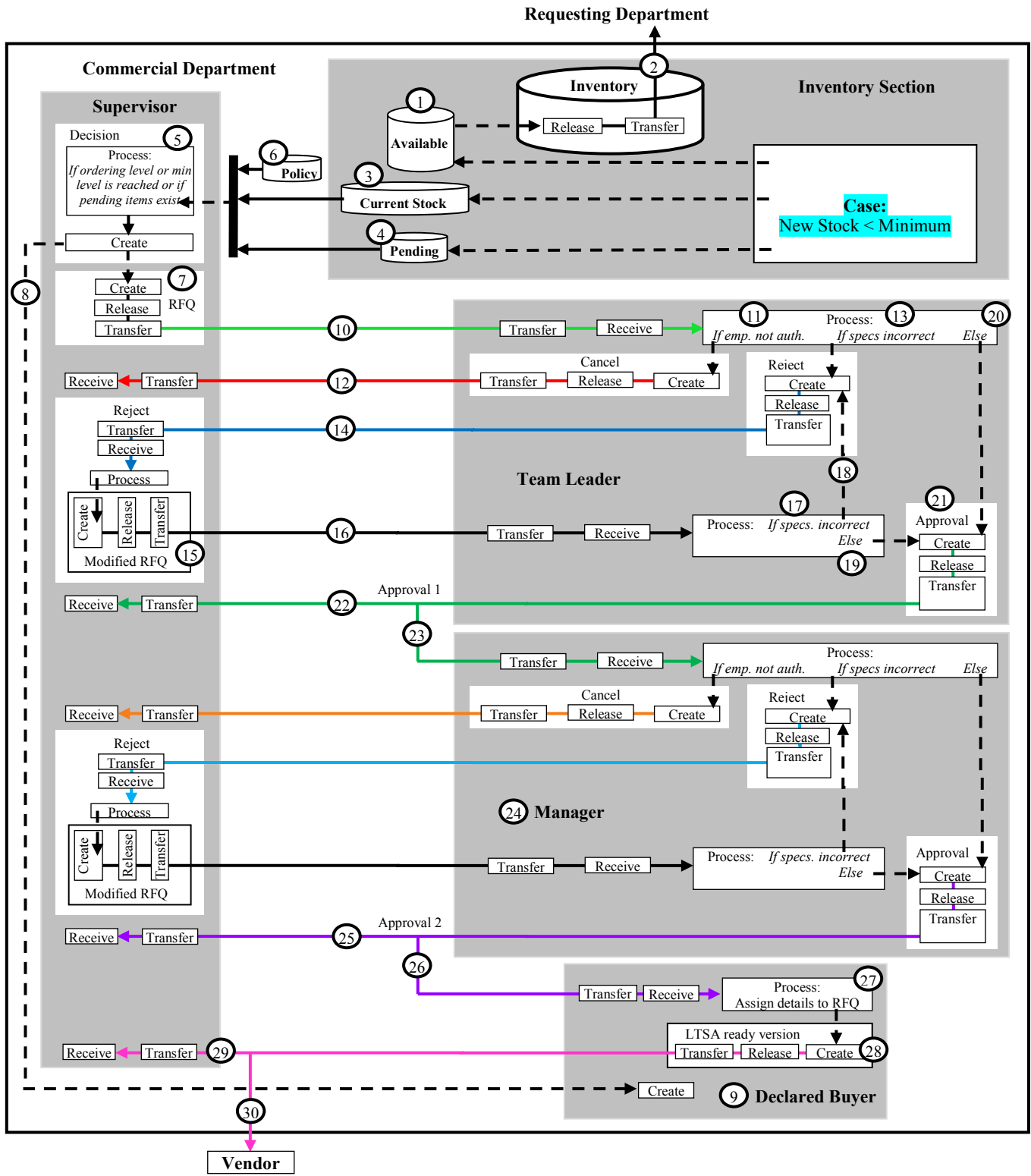

**Fig. 16 TM representation of the case if the new stock for the request is below the minimum**

- The declared buyer is responsible for assigning more details in the RFQ (27), which triggers the creation (28) of a long-term supply agreement (LTSA) (28). The LTSA does not need a bidding process because it is agreed upon with a specific vendor as a single source. It is called long-term because the contract with the vendor states that the price of the requested items shall be fixed for a specific number of years. The LTSA is sent to the supervisor (29 - copy), then the specified vendor (30). The vendor creates his own cycle of preparing and shipping the ordered items according to a specified time limit.

### D. TM representation of receiving the ordered items from the vendor

Fig. 17 shows the TM representation after receiving the ordered items from the vendor. Once the vendor (1) delivers the items (2), the current stock is updated (3). The pending requests (4) in the queue are released (5), one by one, to be processed to extract the requested quantity (6). Each request is processed to determine its quantity. Furthermore, the quantity is processed (7) to release the corresponding number of items in the queue (8). Additionally, the quantity flows to update the current stock (9) and the total pending items (10). Fig. 18 shows the events after receiving the requested items from the vendor. Fig. 19 displays the control over this sequence.

Note that the loop of events for every request in the queue is represented as a second order control over the events required for each request.

The thick horizontal bar at the bottom of Fig. 17 indicates the possible parallelism of $E_5$, $E_6$ and $E_7$. All of these events should end before starting another round of the loop.

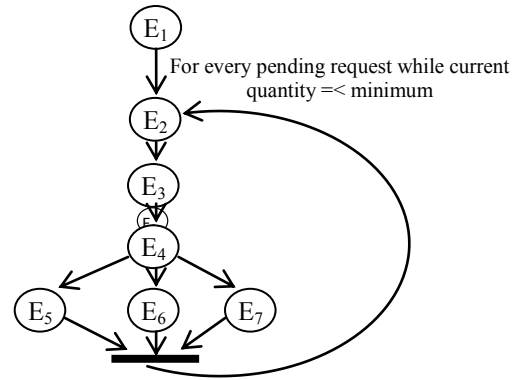

**Fig. 19 The events and their control after receiving the requested items from the vendor**

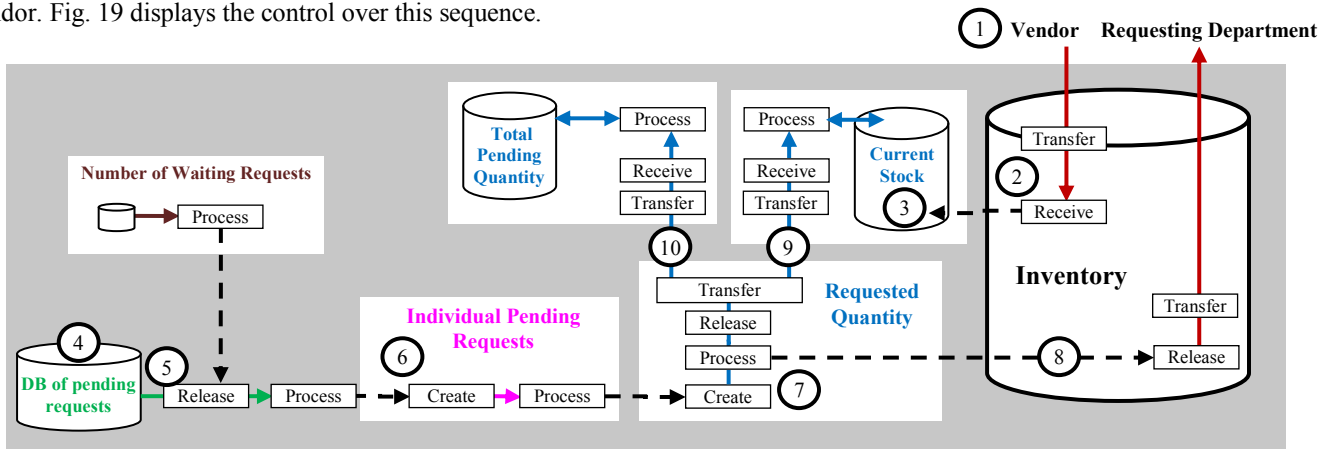

**Fig. 17 TM representation of receiving the requested items from the vendor**

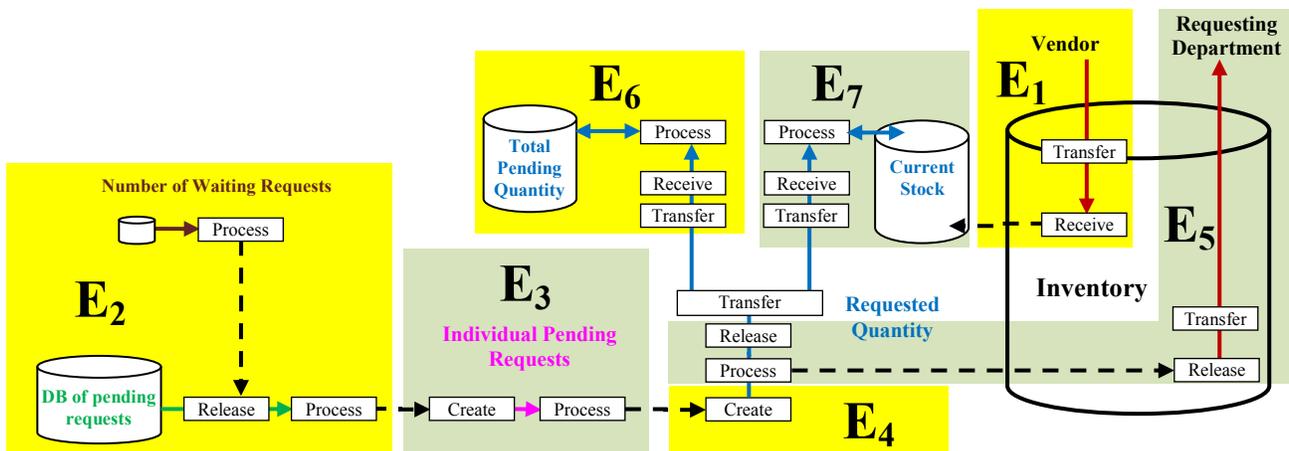

**Fig. 18 The events after receiving the requested items from the vendor**

## V. CONCLUSION

This paper demonstrates the viability of TM modeling for inventory management processes. The resultant conceptual model covers the static, dynamic and control of these processes. This feature of notions' uniformity, based on the simplicity of the TM with its five stages, sets the modeling methodology apart from the heterogeneous diagrammatic representations (e.g., UML) that were displayed in section two of the paper. Further research will establish different potential benefits of the TM approach.